\documentstyle[12pt]{article}
\title{Pion scattering and the skyrmion in $\chi PT$ with many pions}
\author{       {\bf Antonio  DOBADO$^*$}   \\
Department of Physics  \\ Stanford University,
Stanford, CA 94305-4060, USA\\
{\bf  Antonio LOPEZ }\\
 Departamento de F\'{\i}sica Te\'orica \\ Universidad Complutense
de Madrid,
 28040 Madrid, Spain   \\
 and  \\
{\bf John MORALES}    \\
 Departamento de F\'{\i}sica Te\'orica \\
Universidad Auton\'oma de Madrid, 28049 Madrid, Spain}
\begin{document}
\maketitle
\begin{abstract}
The large $N$ effective action of the non-linear sigma model based in the coset
$O(N+1)/O(N)$  is
 obtained. The renormalization of this effective action requires the
introduction of an infinite set of counter terms. However, there exist
particular cases where, at some scale, only a finite number of non-zero
coupling constants are present. This fact make possible  a one parameter fit
of the $I=J=0$ low-energy pion scattering. The corresponding non-local
effective
action is used to study the properties of the  skyrmion
 which is shown to be unstable in this approximation.
\end{abstract}
FT/UCM/14/93, SU-ITP-93-32 \\
{\small *  On leave of absence from
Departamento de F\'\i sica Te\'orica, Universidad Complutense
de Madrid,
 28040 Madrid, Spain}
\newpage
\baselineskip 0.83 true cm
\textheight 20 true cm

\section{Introduction}
{}From the pioneering work by Weinberg [1]
 on phenomenological lagrangians, a lot of work
has been devoted to this field. Specially remarkable was the work
by Gasser and Leutwyler [2]
where this technique was applied, at the one-loop level, to the description of
the
 low-energy behavior of the Nambu-Goldstone bosons (pions)  corresponding to
the
spontaneous breaking of the chiral symmetry of hadron interactions, i.e. the
so-called
Chiral Perturbation Theory ($\chi $PT). More recently,
the phenomenological lagrangian approach has also been used in the context of
the
symmetry breaking sector of Standard Model and, in particular, it was applied
to the scattering of the longitudinal components of the weak gauge bosons [3].

In spite of the great phenomenological success of $\chi $PT, some problems
still
must be solved. Typically  $\chi $PT computations are done to the one-loop
level. This fact puts strong limitations to the region of applicability of
$\chi $PT which is restricted to  the very low-energy regime since, at higher
energies,
conflicts with unitarity appear [4]. Several methods have been proposed in the
literature to improve the unitarity behavior of $\chi $PT like the introduction
of new fields
corresponding to resonances [5], the Pad\'e approximants [4], the inverse
amplitude
 method  [6] etc. More recently it was suggested that it is possible to define
 a sensible
large $N$ approach to $\chi $PT ($N$ being the number of Nambu-Goldstone
bosons)
 which is consistent in principle at any energy.
 In particular, it was shown in [7] how this approach can be used to compute
and renormalize the
pion
scattering amplitude and in [8] it was used to study the $\gamma \gamma
 \rightarrow \pi^0 \pi^0$ reaction. More recently, some of the results of [7]
 were reobtained in [9] by using a different method.

In this work we pursuit this issue further as follows: First we show a
functional
 computation of the effective action for pions in the model
$O(N+1)/O(N)$ to the order of $1/N$ which is non-local. From this effective
action we obtain
the elastic scattering
amplitude for pions. We use this result to make a one parameter fit of the
experimental data corresponding to the $I=J=0$ channel. Then we obtain
the mass functional for the skyrmion from the large $N$ effective action.
 We arrive to the conclusion that the skyrmion is not stable in the many
pion approximation considered here and we finish by discussing which could
be the reasons for that and giving the main conclusions of our work.

\section{The effective action for pions  in the $1/N$
expansion}
 To define the large $N$ limit of $\chi $PT we start from the
two-flavor chiral symmetry group
 $SU(2)_R \times
SU(2)_L$. Then we use the equivalence of the coset spaces  $SU(2)
\times SU(2)/SU(2)=O(4)/O(3)=S^3$ to extend this symmetry pattern to
$O(N+1)/O(N)=S^N$ so that $N$ can be understood as the number of
Nambu-Goldstone bosons (NGB) or pions.
The fields describing these NGB can be chosen as arbitrary coordinates
 on the coset
 manifold $S^N$. The most general $O(N+1)$ invariant lagrangian
 can be obtained as
 a derivative expansion which is covariant respect to both, the space-time and
 the $S^N$ coordinates.
The lowest order is the well known non-linear sigma model lagrangian given by:
\begin{equation}
{\cal L}_{NLSM}=\frac{1}{2}g_{ab}(\pi)
\partial _{\mu}\pi^a\partial^{\mu}\pi^b
 \end{equation}
In our computations we will select the standard coordinates $\pi_a=\pi^a$ where
 $g_{ab}(\pi)=\delta_{ab}+\pi_a\pi_b/(NF^2-\pi^2)$ being $NF^2$ the squared
sphere radius and $\pi^2=
\Sigma_{a=1}^N\pi_a\pi_a$ . The above lagrangian gives rise not only to the
kinetic term but also to interacting terms with arbitrary even number of pions.
These two-derivative interactions produce some difficulties in the quantum
theory
which have been
largely discussed in the literature  [10]. From the path integral point of
view, the proper quantization
prescription is obtained by defining the path integral measure as
$[d \pi \sqrt {g}]$ which is $O(N+1)$ invariant  provided $g$ is the $S^N$
metric determinant.
This factor gives an extra contribution to the non-linear sigma
model lagrangian proportional to $\delta^4(0)$. As it is well known, one can
deal with this new term by using dimensional regularization [11]. In this case
one work in
a $n=4-\epsilon$ dimensional space-time so that the regularized classical
lagrangian
 still is $O(N+1)$ invariant
and, in addition, the measure term can be ignored as a consequence of the rule
$\delta ^n(0)=0$ or equivalently $\int d^nk=0$. In the following we will work
always
in this regularization scheme and hence we will neglect completely the
measure term (for a recent discussion about regularization methods
 in $\chi$PT see [12]).

In order to derive the regularized large $N$ effective action from the above
non-linear sigma model we will extend a technique that has already been used
in the case of the linear theory [13]. First we introduce an auxiliary vector
field
$B^{\mu}$ (in contrast with the $\lambda \Phi ^4$
case where an  auxiliary scalar field has to be introduced).  Then we consider
the new modified lagrangian for the non-linear sigma model:

\begin{eqnarray}
\tilde {\cal L}_{NLSM}&=&\frac{1}{2}\partial _{\mu}\pi_a\partial ^{\mu}\pi^a-
\frac{1}{2}NF^2B_{\mu}B^{\mu}+
B_{\mu}\frac{\pi_a
\partial_{\mu}\pi^a }{\sqrt{1 -\pi^2/NF^2}}
\end{eqnarray}
As there is not any kinetic term for the $B$ field in this lagrangian and it is
quadratic in this field,
a simple gaussian integration makes possible to integrate out this auxiliary
 field in such a way that
the original lagrangian is recovered so that, the pion Green functions obtained
from the lagrangians in eq.1 and eq.2 are the same. However, the new lagrangian
symplifies the computation of  the $N$ factors for any given Feynman diagram.
This is
a consequence of the fact that the lagrangian
in eq.2 does not produce pion self-interactions but $B$ mediated pion
interactions. The
$B$ propagator is just a constant of the order $1/N$ and we have a $2\pi B$
vertex
of the order one, a $4\pi B$ vertex
of the order $1/N$, a $6\pi B$ vertex
of the order $1/N^2$ etc.

Therefore, the most general pion Green function can be obtained by attaching
the external pions in all possible ways to the $B$ legs of the most general $B$
Green function
(see Fig 1.a for an example). The most general connected $B$ Green function
(represented in Fig.1 by
 a black blob) will contain loops both of pions and $B$'s and in general can be
written in terms of
$B$ Green functions containing only pions loops (represented in Fig.1 by a
dashed
 blob)
attached to $B$ lines forming loops (see Fig.1b). In addition, it is not
 difficult to realize that
the dashed $B$ functions are as much of the order of $N$ (see Fig.1c).
Therefore,
taking into account
that the $B$ propagators are of the order of $1/N$ we conclude that, to
leading order in the $1/N$ expansion, black $B$ Green functions are
equal to the dashed $B$ Green functions or, in other words, to leading order we
do not have to consider
diagrams with $B$ loops (see Fig.1d).

Thus in the large $N$ limit we just need to compute the generating
functional for the 1PI $B$ Green functions and then  couple the pions to
the external $B$ legs at the tree level. Moreover, we are interested only in
the two point and the four point pion Green functions which are of the order of
$1$ and $1/N$ respectively. This means that we need to consider only one point
and two point $B$ functions. In this case it is not difficult to see that the
only vertex that must be considered is $2\pi B$ since the others necessarily
produce
pion loops with only one vertex. However this kind of loop vanishes because our
pions are massless and we are using dimensional regularization. In principle,
 there could
 be also a contribution of the order of $1/N$
to the two-pion function coming from the one leg $B$ diagram with one
 $B$ loop inside. Nevertheless,
this diagram is proportional to the external $B$ momentum so that it does not
contribute to the two-pion function.

Thus, the $B$ 1PI generating functional (or $B$ effective action) that we
have to compute can be written as:
\begin{equation}
 i\Gamma_{eff} [B]=i\int d^nx(-\frac{1}{2}NF^2B^2)+\log
\int[d\pi]\exp{i\tilde S[\pi,B]}
\end{equation}
where:
\begin{equation}
\tilde S[\pi,B]=
\int d^nx (-\frac{1}{2}\pi_a\Box \pi^a+B_{\mu}\pi_a
\partial^{\mu}\pi^a)
\end{equation}
As the  functional integral is quadratic in the pion field it can be
computed exactly
by the use of standard methods. We introduce the operator:
\begin{eqnarray}
K_{ab}(x,y)  &=&\frac{-\delta^2 \tilde S}{\delta \pi^a(x)\delta\pi^b(y)}
\mid_{\pi= \bar{\pi}} \\   \nonumber
&=&
\delta_{ab}\Box_x\delta(x-y)+\delta_{ab}\partial^x_{\mu}B^{\mu}
(x)\delta(x-y) \\   \nonumber
          &=&\Box_{ab}(x,y)+\Delta_{ab}(x,y)
\end{eqnarray}
where $\bar{\pi}$ is the solution of the equation of motion
$(\partial_{\mu}\partial^{\mu}+\partial_{\mu}B^{\mu})\bar{\pi}=0$
and therefore we can write:
\begin{equation}
\Gamma_{eff} [B]=\int d^nx(-\frac{1}{2}NF^2B^2)+
\tilde S[\bar \pi,B]+\frac{i}{2}Tr\log K
\end{equation}
As usual the trace of  the $\log K$ operator is expanded as:
\begin{eqnarray}
Tr\log K  &=&Tr\log \Box+Tr\log (1+\Box^{-1}\Delta) \\  \nonumber
&=&
Tr\log \Box+\sum_{k=1}^{\infty}
\frac{(-)^{k+1}}{k}Tr(\Box^{-1}\Delta)^k
\end{eqnarray}
As it was mentioned before, we only need to compute  the first
two terms of this sum corresponding  to the one and
two points $B$ functions respectively. In addition, the one point function
vanishes
 since it is proportional to a tadpole integral.
Therefore, the only contribution to the $B$ effective action is:
\begin{equation}
Tr(\Box^{-1}\Delta)^2=N\int
d^nxd^nyB_{\mu}(x)B_{\nu}(y)\int\frac{d^nk}{(2\pi)^n}
k^{\mu}k^{\nu}e^{ik(x-y)}\int\frac{d^nq}{(2\pi)^n}\frac{1}{(k+q)^2q^2}
\end{equation}
so that, by neglecting the no $B$-leg contribution and terms with more than
two legs we find:
\begin{equation}
\Gamma_{eff}[B]=-\frac{1}{2}\int d^nxd^nyB_{\mu}(x)
\Gamma^{\mu\nu}(x,y)B_{\nu}(y)
\end{equation}
where:
\begin{equation}
\Gamma_{\mu\nu}(x,y)=NF^2g_{\mu\nu}\delta(x-y)-
\frac{N}{2}\int\frac{d^4k}{(2\pi)^4}e^{ik(x-y)
}\frac{k_{\mu}k_{\nu}}{(4\pi)^2}(N_{\epsilon}
+2+\log\frac{\mu^2}{-k^2})
\end{equation}
and as usual $N_{\epsilon}=2/\epsilon+\log 4\pi-\gamma$ and $\mu$ is an
arbitrary dimensional scale.
Next we add to this action the coupling with the external pions
$B_{\mu}\pi_a\partial^{\mu}\pi^a$:
\begin{equation}
\Gamma_{eff}[\pi,B]=\int d^4x B_{\mu}\pi_a\partial^{\mu}\pi^a+\Gamma_{eff}[B]
\end{equation}
Now we use this action to obtain  the $B$ field  as a functional
of the $\pi$ field through the corresponding equation of motion:
\begin{equation}
\frac{\delta \Gamma_{eff}}{\delta B^{\mu}}=\pi_a\partial_{\mu}\pi^a-\int d^4y
\Gamma^{\mu\nu}(x,y)B_{\nu}(y)=0
\end{equation}
the solution to this equation can be written as:
\begin{equation}
\bar B_{\mu}(x)=\frac{1}{2}\int d^4y
\Gamma^{-1}_{\mu\nu}(x,y)\partial^{\nu}\pi^2(y)
\end{equation}
The effective action for pions can  be written as:
\begin{equation}
S_{eff}[\pi]=\int d^4x(\frac{1}{2}\partial_{\mu}\pi_a\partial^{\mu}\pi^a
+\pi_a\partial^{\mu}\pi^a\bar B_{\mu}
-\frac{1}{2}\int d^4y\bar B_{\mu}(x)\Gamma^{\mu\nu}(x,y)\bar  B_{\nu}(y)
)
\end{equation}
by computing the inverse of the $\Gamma_{\mu\nu}(x,y)$ operator
 appearing in eq.10  we finally obtain:
\begin{eqnarray}
S_{eff}[\pi]&=&\int d^4x \frac{1}{2}\partial_{\mu}\pi_a\partial^{\mu}\pi^a  \\
\nonumber
&+&\frac{1}{8NF^2}\int d^4x d^4y \pi^2(x)\pi^2(y)
\int
\frac{d^4q}{(2\pi)^4}e^{-iq(x-y)}\frac{q^2}{1-\frac{q^2}{2(4\pi)^2F^2}(N_{\epsilon}
+2+\log\frac{\mu^2}{-q^2})}  \\ \nonumber
 &+& O(1/N^2)
\end{eqnarray}
 Here the denominator
 must be understood as the corresponding  formal geometric series.

\section{Renormalization of the effective action}

Once the regularized effective action for pions has been obtained in
the large $N$ limit we have to deal with the renormalization of this result.
 As the
 kinetic term of this effective action is just the classical one,
 no renormalization of the pion wave function is needed i.e. $Z_{\pi}=1$.
However,
in the effective action we find four pion interaction terms which are divergent
and proportional to
 arbitrary high even powers of the momenta. This of course reveals the well
known
fact that the non-linear sigma model is not renormalizable in the standard
sense.
Nevertheless, we can absorb all the divergences by adding to the original
lagrangian
in eq.1 an infinite set of new counter terms that can be chosen for example as:
\begin{equation}
{\cal L}^{ct}=
\sum_{k=1}^{\infty}\frac{g_k}{8N}\pi^2(\frac{-\Box}{F^2})^{k+1}\pi^2+O(1/N^2)
\end{equation}
Apart from these, on could also in priciple to introduce many other
terms. However, here we will
adopt the philosophy of introducing only a  minimal set of  terms just
 to absorb the divergences appearing
in the non-linear sigma model to leading order in the $1/N$ expansion.

Now, following similar steps as in [7], it is straightforward to obtain
the renormalized pion effective action:
\begin{eqnarray}
S_{eff}[\pi] &=&\int d^4x \frac{1}{2}
\partial_{\mu}\pi_a\partial^{\mu}\pi^a  \\ \nonumber
 &+&\frac{1}{8NF^2}\int d^4x d^4y \pi^2(x)\pi^2(y)
\int \frac{d^4q}{(2\pi)^4}e^{-iq(x-y)}\frac{q^2G^R(q^2;\mu)}{1-\frac{q^2}
{2(4\pi)^2F^2}G^R(q^2;\mu)
\log\frac{\mu^2}{-q^2}}  \\ \nonumber
 &+& O(1/N^2)
\end{eqnarray}
where $G^R(q^2;\mu)$ is the generating function for the renormalized coupling
constants $g_k(\mu)$ at the scale $\mu$:
\begin{equation}
G^R(q^2;\mu)=\sum_{k=0}^{\infty}g_k(\mu)(\frac{q^2}{F^2})^k
\end{equation}
where by definition $g_0(\mu)=1$ for every $\mu$. Since $Z_{\pi}=1$, the
renormalization
group equation followed by the pion effective action is:
\begin{equation}
(\frac{\partial}{\partial \log \mu}+\sum_{k=0}^{\infty}\beta_k
\frac{\partial}{\partial g_k(\mu)})S_{eff}[\pi]=0
\end{equation}
where $\beta_k=dg_k(\mu)/d\log \mu$. From this equation we can obtain very
easily the
renormalization group equation for $G^R(q^2;\mu)$ which can be integrated
to give the evolution equation:
\begin{equation}
G^R(q^2;\mu)=\frac{G^R(q^2;\mu_0)}{1+\frac{q^2}{2(4\pi)^2F^2}G^R(q^2;\mu_0)
\log(\frac{\mu^2}
{\mu_0^2})}
\end{equation}
By using this equation and eq.18 it is very easy to get the evolution
 equations for all the renormalized coupling constants $g_k(\mu)$.
Again because $Z_{\pi}=1$
and the fact that pions continue being massless we can read very easily the
elastic pion
scattering amplitude from eq.17 recovering the result of [7] (one missing
factor $2$ in this reference has been
 corrected here):
\begin{equation}
T_{abcd}(s,t,u)=A(s)\delta_{ab}\delta_{cd}+A(t)\delta_{ac}\delta_{bd}+
A(u)\delta_{ad}\delta_{bc}
\end{equation}
where:
\begin{eqnarray}
A(s)&=&\frac{s}{NF^2}\frac{G^R(s;\mu)}{1-\frac{s}{2(4\pi)^2F^2}G^R(s;\mu)
\log(\frac{\mu^2}{-s})}+ O(1/N^2)
\end{eqnarray}
Note that at low energies this $A(s)$ function behaves as $s/NF^2$ and this
fact
is important for two reasons. First, we realize that it is compatible with
the low energy theorems which state that it must go in this limit as
 $s/f^2_{\pi}$ (in the chiral limit that we are considering here). Second
we can use this low-energy theorems to set the constant $F$ in terms of the
pion decay
 constant $f_{\pi}\simeq 96 MeV$ as $f^2_{\pi}=NF^2$ with $N=3$.

In addition to this scattering amplitude, there is another important
object that can be obtained from the effective action in eq.17. For  static
 pion field configurations $\pi_a(\vec x)$ the energy density is given
by minus the effective lagrangian i.e. $-{\cal L}_{eff}(\vec x)$ which can
be read off from eq.17. Therefore, the mass functional or, equivalently,
the total energy of the configuration can be written as:
\begin{eqnarray}
M[\pi]=\int d\vec x(\frac{1}{2}\partial_i \pi_a \partial^i \pi^a
+\frac{1}{8NF^2}\pi^2[\frac{-G^R(\nabla;
\mu)\nabla}
{1-\frac{1}{2(4\pi)^2F^2}G^R(\nabla;\mu)
\log(\frac{\mu^2}{-\nabla})\nabla}](\pi^2)
\end{eqnarray}
where the meaning of this non-local mass functional is
given by the appropriate Fourier transform.

In order to use eq.22 or eq.23 to do phenomenology we need to define the
generating function $G^R(q^2;\mu)$ or, in other words, the infinite number
of renormalized couplings constants $g_k(\mu)$. A typical example is
 the linear sigma model. In this case,
 the appropriate choice is:
\begin{eqnarray}
g_k(\mu)&=&(\frac{1}{2\lambda(\mu)})^k        \nonumber   \\
G^R(s;\mu)&=&
\frac{1}{1-\frac{s}{2\lambda(\mu)F^2}}
\end{eqnarray}
where $\lambda(\mu)$ is the renormalized coupling constant of the  linear
sigma model computed in the large $N$ limit [14]. With this definition, eq.22
and eq.20
provide the corresponding
well known scattering amplitude and evolution equation for $\lambda$. In
addition,
the amplitude in eq.22 has a pole in the second Riemann sheet
that has to be understood as the physical scalar resonance  appearing in the
linear sigma model i.e. the Higgs resonance. (see [14] and [15] for
a discussion on this point).

However, the $O(N+1)$ linear sigma model is not the theory that reproduces the
hadron
interactions at low energies. In principle, there are two possible approaches
for the
 determination of the renormalized coupling constants appearing in the pion
 effective action of eq.17 . From the theoretical point of view one could try
to
compute these parameters from a more fundamental theory of strong interactions
like
QCD. From the phenomenological point of view, one can try to fit these
constants
from the experiment, for example the low-energy pion scattering data. As the
theoretical
computation seems to be quite involved, it is worth to make the
phenomenological
determination of the couplings first in order to test the quality of the
whole $N$ approach to this case.

\section{Fitting the elastic pion scattering}

 Using  eq.22 to fit the experimental pion scattering data requires
to face the fact of having an infinite number of parameters i.e. the scale
$\mu$ and the values of the renormalized coupling constants $g_k(\mu)$
at the scale $\mu$. However one can solve this problem in a easy way just
by considering only particular cases where all the coupling
constants but a finite set $g_1$, $g_2 \dots g_r$ vanish at some scale $\mu$.
These models are defined just by a finite number of parameters ($\mu$
 and the $r$ coupling constants renormalized at this scale) and therefore can
be used to fit the experimental data. In particular, one can consider the
extremal
case of having all the renormalized couplings equal to zero at some  scale
$\mu$ i.e.
$g_k(\mu)=0$ for all $k>0$. The
model so obtained has only one parameter (the $\mu$ scale)
and it can be considered in some sense as the non-linear sigma model
renormalized at the
leading order of the $1/N$ expansion.

 Of course, it could happen that
the underlying theory (say QCD) were not of this type. However this would not
be
 a serious problem at very low energies because, for natural
 values of the renormalized coupling constants and the renormalization scale
$\mu$,
only the very first terms
of the generating function $G^R(q^2;\mu)$ series in eq.18 contribute to the
effective
action and the scattering amplitude. However, at higher energies the effect of
the
infinite  set of couplings constants, when different from zero, could be
important. For example,
 they can conspire to give rise to
 resonances which appear as poles in
the second Riemann sheet of the partial waves amplitudes as it happens in the
linear sigma model. Of course, in this case the value of the couplings can also
be obtained
from the underlying theory or by making extra assumptions on the behavior of
the amplitudes.

For comparison with the experiment we need also to compute the phase shifts for
the
different channels. These are obtained from the corresponding partial waves
 amplitudes. The proper generalization of the standard $SU(2)$ isospin
projections
to the $O(N)$ case is given by [9]:
\begin{eqnarray}
T_0=NA(s)+A(t)+A(u)  \\   \nonumber
T_1=A(t)-A(u) \\     \nonumber
T_2=A(t)+A(u)
\end{eqnarray}
and the partial waves are given by:
\begin{equation}
a_{IJ}=\frac{1}{64\pi}\int_{-1}^1T_I(s,cos \theta)P_J(cos \theta)d(cos \theta)
\end{equation}
Then, the $a_{00}$ amplitude is of the order of one but, for instance, the
$a_{11}$ and the $a_{20}$ amplitudes are of the order of $1/N$. Thus at the
leading order of the large $N$ expansion $a_{00}(s)=NA(s)/32 \pi+O(1/N)$,
$a_{11}(s)=0+O(1/N)$ and $a_{20}(s)=0+O(1/N)$ (remind that $A(s)$ is of the
order
of $1/N$). Now it is easy to show that in this approximation
 the partial waves have the proper cut structure and also
 are elastic unitary i.e. $Ima_{00}=\mid a_{00}\mid ^2+O(1/N) $ just above the
unitarity cut
as they must be.

In fig.2 we show the result of our fit of the $I=J= 0$
 phase shift channel for elastic pion scattering. This fit has been
 done with only one parameter, or in other words, by using the very particular
model
where all the renormalized coupling constants are zero at some scale $\mu$. As
discussed above,
the only parameter in this specially
 simple case is the
scale $\mu$. The fit in fig.2 corresponds to a value of $\mu^2=600000 MeV^2
(\mu\simeq 775 MeV)$. For other
 channels like $I=J=1$ or $I=2, J=0$ the prediction of the leading order of the
large $N$ expansion is that there is no interaction. This could appear as a
rather poor prediction but it is not the case. By looking at the experimental
data one can realize that the phase shifts in these channels grow very slowly
with
energy, for instance, at $500 MeV$ of center of mass energy we have
$\delta_{11}
\simeq 5 $ degrees
  and $\delta_{20}\simeq -7 $ degrees whilst $\delta_{00}\simeq 38
 $ degrees at the same energy. The conclusion is that for energies below, let
say, $500 MeV$
our simple one parameter fit gives rise to errors in the phase shifts which are
 lesser than $10$ per cent (refered to the $I=J=0$ channel)
for the three channels considered. Of course, for larger energies things become
completely different particularly in the $I=J=1$ channel where the onset of
 the $\rho$
resonance makes this channel to be strongly interacting. One is then tempted to
make an interpretation  of the scale $\mu$ as some kind of cutoff signaling
the range of applicability of our approach and, in fact,  this interpretation
was discussed
in [7]. From this point of view, it is quite interesting to realize how close
is the $\mu$
 fitted value $(\mu\simeq 775 MeV)$ to the $\rho$ mass. In some way, by fitting
the $I=J=0$
channel, the model is telling us where new things can appear in other channels
like the $I=J=1$
(the $\rho$ resonance) and thus setting the limits of applicability of the
model.
However, we would like to stress that our renormalization method is
 completely
 consistent and our results are formally valid at any energy independently of
the goodness of the
fits we can obtain with them.

\section{The skyrmion in the large $N$ limit}

As it was mentioned above, the second phenomenological application of the large
$N$ expansion that will be considered here is the description of the nucleon
 in the context of the so-called skyrmion model. As it is well known,
the skyrmion [21] is a topological
soliton of the $SU(2)_L\times SU(2)_R/SU(2)_{L+R}$ chiral model. Finite
 energy and static pion field
configurations are  continous maps from the compactified space
$S^3$ to the coset space $S^3$. These applications can be classified in
homotopically non-equivalent classes since $\pi_3(S^3)={\it Z}$. Therefore
every map can be labeled by an integer number. In the skyrmion model
of the nucleon this topological number is interpreted as the baryonic number
of the field. The skyrmion model has proved to give a qualitative description
of many of the nucleon properties [22] including (in an highly non-trivial way)
the spin [23].   The standard ansatz for the
Skyrmion field is $U(\vec{x}) = exp(i f(r) \hat{x}_a \sigma_a)$ with
$f(0) = \pi$ and $f(\infty) = 0$ being the right boundary conditions for the
chiral angle $f(r)$ in order to have baryonic number equal to one. In this
description of the skyrmion the chiral field is taken to be an element
of $SU(2)$ which is equivalent to our coset space $S^3$. In terms of the
coordinates used here, the skyrmion ansatz is given by $\pi^a(\vec x)
=\sqrt{N}F\sin f(r)\hat{x}^a$. In order to obtain the mass of the skyrmion in
the approximation considered in this work we have to introduce this anzatz
in the mass functional in eq.22 with the renormalized coupling constants
obtained
 from the fit of the elastic pion scattering
described in the previous section. Then, it is not very difficult to find:
\begin{eqnarray}
M[f]&=&NF^22\pi\int_0^{\infty}dr r^2
(2 \frac{\sin^2f(r)}{ r^2}+\cos^2 f(r) f'(r)^2) \\ \nonumber
&+&NF^2\int_0^{\infty}dr r \sin ^2f(r)\int_0^{\infty}dr' r' \sin ^2f(r')
\int_0^{\infty}dk\sin(kr)\sin(kr')   \\ \nonumber
& &\frac{k^2G^R(-k^2;\mu)}{1+\frac{k^2}{2(4\pi)^2F^2
}G^R(-k^2;\mu)\log\frac{\mu^2}{k^2}}
\end{eqnarray}
Now the mass functional depends only on the chiral angle $f(r)$. By minimizing
this functional we can find in principle the mass and the shape of the skyrmion
in
this large $N$ approximation. In order to make this task easier we can use
some reasonable parametrization of the chiral angle such as
the one by Atiyah and Manton [24]:
\begin{equation}
f_{R}(r) = \pi ( 1 - ( 1 + R^2/
r^2 )^{-1/2} )
\end{equation}
where the parameter $R$ is a measure of the size of the Skyrmion.
In fig.3 we show the results obtained when this parametrization
is inserted in eq.26 with the $G^R(k^2;\mu)$ obtained in the previous
section i. e. we follow the philosophy of  [25]
where the authors use the skyrmion model to predict the
mass of the proton from the pion scattering data. The first and
obvious conclusion from fig.3 is that, at least for this parametrization,
 the mass functional has not any minimum but the trivial one at $R=0$
so that the skyrmion tends to shrink to zero size in this approximation. In
fact we have checked that this is also the case when other parametrizations
of the chiral angle are used.

 Therefore, everything seems to indicate that in the  large $N$ limit
considered here,
the skyrmion is not stable i.e. we are in a similar situation  to the one
found when one works with the two derivatives term in eq.1. This is a bit
disappointing since one of the most appealing properties of the large $N$
approximation is that it includes the effect of higher derivative terms and
the corresponding higher loops contributions. This fact improves the unitarity
behavior of the amplitudes and one could think that it could also
improve the skyrmion description. However, this is not the case, at least in
the
leading order considered here. A simple way to
understand why this is not so is the following. In the standard approach to the
skyrmion
one typically adds to the two derivatives term in the lagrangian
one four derivative term  called the skyrme term. This term produce an
stable skyrmion in such a way that the contribution to the total energy of
the skyrmion coming from the two terms is exactly the same. It is also possible
 to
add another four derivative term called the non-skyrme term but its
 contribution can turn the skyrmion unstable. One-loop effects
can also be included in the mass functional [26]. One possible
 interpretation of the origin of the
skyrmion and the non-skyrmion terms is to understand them
as the low energy remnant of the tail
of one vector $(I=J=1)$ and  one scalar $(I=J=0)$ resonance respectively. In
fact it is not difficult to relate the skyrme and the
non-skyrme parameters with the masses of the corresponding resonances
(see for instance [4]). Then,
the consequence
  of this picture
is that the existence of one vector resonance like the $\rho$ produces an
stable
skyrmion (and in fact it is possible to relate the $\rho$ and the nucleon
masses)
but one scalar resonance do not produce an stable skyrmion by itself
and even it can make unstable the skyrmion when the vector resonance is also
present.

If this simple picture were correct (as it seems to be) the $\rho$ resonance
would be the responsible of the stability  of the nucleon in the skyrmion
 model (or at least of the pion cloud). Then it would be possible to
understand why the skyrmion is not stable in the large $N$ limit approximation
considered in this paper. As we saw in the previous section this approach can
describe
well the elastic pion scattering in the $I=J=0$ channel  but it does not
 reproduce
the $\rho$ resonance (yet in some way it predicts its existence) and therefore
it does
 not contain the piece that seems to keep the
skyrmion stable in other much more simple approach that do not include
higher derivatives neither higher loops effects.

\section{Conclusions}
We have computed the effective action of the $O(N+1)/O(N)$ non-linear sigma
model to leading order
on the $1/N$ expansion. This action is divergent but it can be properly
renormalized by introducing
an infinite set of counter terms in a systematically way. The energy dependence
of the corresponding renomalized couplings
can be obtained thus providing a very nice example of how $\chi PT$ can work
beyond the one-loop approximation.
With an appropriate choice of these renormalized couplings we can reproduce
particular models with the same
symmetry breaking pattern as, for example, the linear sigma model.

We can consider also models with a finite
number of couplings which can be used to fit the elastic pion scattering
experimental data since the
partial waves amplitudes have a very good unitarity behavior. Then, with only
one
parameter it is possible to reproduce well the phase shifts for energies below
$500 MeV$. This parameter can
be also interpreted as some kind of cutoff setting the limits of the model and,
in fact, the fitted value is
basically the $\rho$ mass which in principle cannot be reproduced since, at
leading order in  the $1/N$ expansion,
the $I=J=1$ channel is not interacting.

 We have also studied the mass functional of the skyrmion solutions and we
arrived to the
conclusion that the skyrmion solution is not stable at the leading order in the
$1/N$ expansion. This fact is probably related
with the above mentioned absence of interaction in the $I=J=1$ since in other
approaches the $\rho$ resonance
stabilizes the skyrmion.

In any case, we consider the $1/N$ expansion of $\chi PT$ as a very interesting
 alternative complementary
 to the standard one-loop computations. Of course, much more work is needed
 like including pion mass effects, going to the
next to leading order in $1/N$, considering other ways to take the limit like
the one based in the
 $SU(N)_L\times SU(N)_R/SU(N)_{L+R}$ non-linear sigma model and finally
 (and this was our original  motivation)
the applications to the description of the symmetry breaking sector of the
Standard Model as a way to parametrize
the future Large Hadron Collider (LHC) data. Work is in progress in all these
directions.

\section{Acknowledgments}
This work has been supported in part by the Ministerio de Educaci\'on y Ciencia
(Spain) (CICYT AEN90-0034) and COLCIENCIAS (Colombia). A. D. thanks also
 the Gregorio del Amo foundation
for support and S. Dimopoulos and the Department of Physics of
 Stanford University for their kind hospitality.

\newpage
\begin{center}
{\large \bf Figure Captions}
\end{center}
{\bf Figure 1:}a)  A typical eight pion legs Feynman diagram  where the  three
$B$ legs function appears (pion lines are narrow and $B$ lines are broader) b)
The complete
 connected three $B$ legs
 function (black blob) in terms of the $B$ functions containing only
pion loops (dashed blobs) c) Some diagrams contributing to the $B$ functions
containing only pion loops d) Relation between the complete $B$ functions
 and the $B$ functions containing only pion loops at the leading order
in the $1/N$ expansion.

{\bf Figure 2:} Phase shifts of the elastic pion scattering for the $I=J=0$
channel. The
 continuous line represents the one parameter fit described in the text. The
 experimental data
corresponds to: $\bigtriangleup$ ref.[16],
$\bigcirc$ ref.[17],$\Box$ ref.[18],$\diamondsuit$ ref.[19],
$\bigtriangledown$ ref.[20].

{\bf Figure 3:} Mass of the skyrmion in the Large $N$ limit for the Atiyah and
Manton
parametrization in terms of the $R$ parameter. Dotted line is the contribution
of the kinetic term, dashed line is the interaction energy and the full line is
the
total mass of the skyrmion.

\end{document}